\documentclass[12pt]{article}
\usepackage{graphicx}
\usepackage{amsmath}
\usepackage{amssymb}
\usepackage{amsfonts}
\usepackage{amsthm}
\usepackage{bm}
\usepackage{verbatim}
\usepackage{color}
\usepackage{xcolor}
\usepackage[utf8]{inputenc}
\usepackage{hyperref}
\usepackage{overpic}

\definecolor{mydarkblue}{RGB}{77, 89, 161}
\definecolor{myred}{RGB}{191, 78, 69}
\definecolor{mygreen}{RGB}{100, 148, 100}
\definecolor{myviolet}{RGB}{105, 85, 145}
\definecolor{myorange}{RGB}{228, 152, 78}
\definecolor{myblue}{RGB}{79, 169, 209}

\newcommand{\set}[2]{\{ \, #1 \mid #2 \, \}}
\newcommand{\setbig}[2]{\big\{ \: #1 \;\big|\; #2 \: \big\}}

\renewcommand{\epsilon}{\varepsilon}

\newcommand{\Image}{\mathop{\mathrm{Im}}}
\newcommand{\Dom}{\mathop{\mathrm{Dom}}}

\newtheorem{theorem}{Theorem}

\newtheorem{lemma}{Lemma}
\newtheorem{claim}{Claim}
\newtheorem{definition}{Definition}
\newtheorem{example}{Example}

\begin{document}

\title{A hierarchy of reversible finite automata}
\author{Maria Radionova\footnote{%
Saint Petersburg University, 7/9 Universitetskaya nab., St. Petersburg, 199034 Russia}
\and Alexander Okhotin\footnote{%
Department of Mathematics and Computer Science, St.~Petersburg State University, 14th Line V. O., 29, Saint Petersburg 199178, Russia}
}

\maketitle
\sloppy

\begin{abstract}
In this paper, different variants of reversible finite automata are compared, and their hierarchy by the expressive power is established. It is shown that one-way reversible automata with multiple initial states (MRFA) recognize strictly more languages than sweeping reversible automata (sRFA), which are in turn stronger than one-way reversible automata with a single initial state (1RFA). The latter recognize strictly more languages than one-way permutation automata (1PerFA). It is also shown that the hierarchy of sRFA by the number of passes over the input string collapses: it turns out that three passes are always enough. On the other hand, MRFA form a hierarchy by the number of initial states: their subclass with at most $k$ initial states (MRFA$^k$) recognize strictly fewer languages than MRFA$^{k + 1}$, and also MRFA$^k$ are incomparable with sRFA. In the unary case, sRFA, MRFA$^k$ and MRFA become equal in their expressive power, and the inclusion of 1RFA into sRFA remains proper.
\end{abstract}

\tableofcontents

\section{Introduction}

Classical one-way deterministic automata (1DFA) have an interesting variant: \emph{reversible automata} with one or multiple initial states (1RFA, MRFA), which were first studied by Angluin~\cite{Angluin} and by Pin~\cite{Pin1987}. Such automata model reversible computation, which is important to study, because reversibility is the only potential way to create computer hardware which does not heat up during work. Indeed, according to Landauer's principle~\cite{Landauer} irreversible erasure of one bit of information causes generation of a certain amount of heat, and this physical limit can be overcome only by logically reversible devices. Among recent studies of 1RFA, there is a paper by H\'eam~\cite{Heam} on the relative succinctness of 1DFA and 1RFA. Holzer and Jakobi~\cite{HolzerJakobi} obtained results about computational complexity of minimization and hyper-minimization problems for 1RFA. The minimal reversible automata were also studied by Holzer et al.~\cite{HolzerJakobiKutrib}. Other decision problems for reversible automata were investigated by Birget et al.~\cite{Birget}.

Another model of reversible automata studied in the literature are the \emph{one-way permutation automata} (1PerFA), defined by an additional restriction: unlike reversible automata, 1PerFA always read input data until the end, that is, their transition function is fully defined and hence forms a permutation on the set of states. One-way permutation automata were first studied by Thierrin~\cite{Thierrin}, and a family of languages recognized by 1PerFA is called the group languages, because the syntactic monoid of such automata is a group. Recently, Jecker, Mazzocchi and Wolf~\cite{JeckerMazzocchiWolf} investigated the computational complexity of the decomposition problem for permutation automata. Also, Hospod\'ar and Mlyn\'ar\v{c}ik~\cite{HospodarMlynarcik} studied the state complexity of operations on permutation automata, while Rauch and Holzer~\cite{RauchHolzer} obtained the complexity of operations on permutation automata in the number of accepting states.

Models of automata can be compared in their expressive power, that is, by inclusion of the families of languages recognized by different models. One-way reversible automata recognize strictly fewer languages than classical deterministic automata: for example, they cannot recognize a regular language $a^*b^*$. At the same time, permutation automata turn out to be an even more constrained model: in particular, they cannot recognize any finite language, whereas reversible automata recognize all finite languages. In this paper, this hierarchy is completed by other variants of deterministic, reversible and permutation automata.

Among the models considered, there are also two-way reversible and permutation automata. Classical two-way deterministic automata (2DFA) can return to previously read symbols of an input string while processing it. The possibility to read a string in both directions does not allow 2DFA to recognize more languages than one-way deterministic automata. But as soon as reversible automata are allowed to read a string in both directions, they can recognize all regular languages, as was shown by Kondacs and Watrous~\cite{KondacsWatrous}, which exceeds the expressive power of one-way reversible automata (1RFA). Between one-way and two-way automata, an intermediate model of \emph{sweeping automata}~\cite{Sipser} was studied. These automata alternate between reading the input string from left to right and from right to left. Recently, a sweeping variant of permutation automata (2PerFA) was studied by the authors~\cite{RadionovaOkhotin}, and it was shown that the new model recognizes exactly the same languages as 1PerFA.

Another variant of automata is that with multiple initial states: these are one-way automata that can choose non-deterministically one of their initial states before reading an input string. Deterministic automata with multiple initial states (MDFA)~\cite{HolzerSalomaaYu} recognize exactly the family of regular languages, that is, their expressive power is the same as 1DFA's. A similar result holds for permutation automata: permutation automata with multiple initial states (MPerFA) cannot recognize more languages than 1PerFA, and this directly follows from the closure of the group languages under the union operation. In the case of reversible automata, as a matter of fact, Pin~\cite{Pin1987} studied the model with multiple initial states (MRFA).

\begin{figure}[t]
    \centerline{%
    \begin{overpic}[scale=0.4,permil]{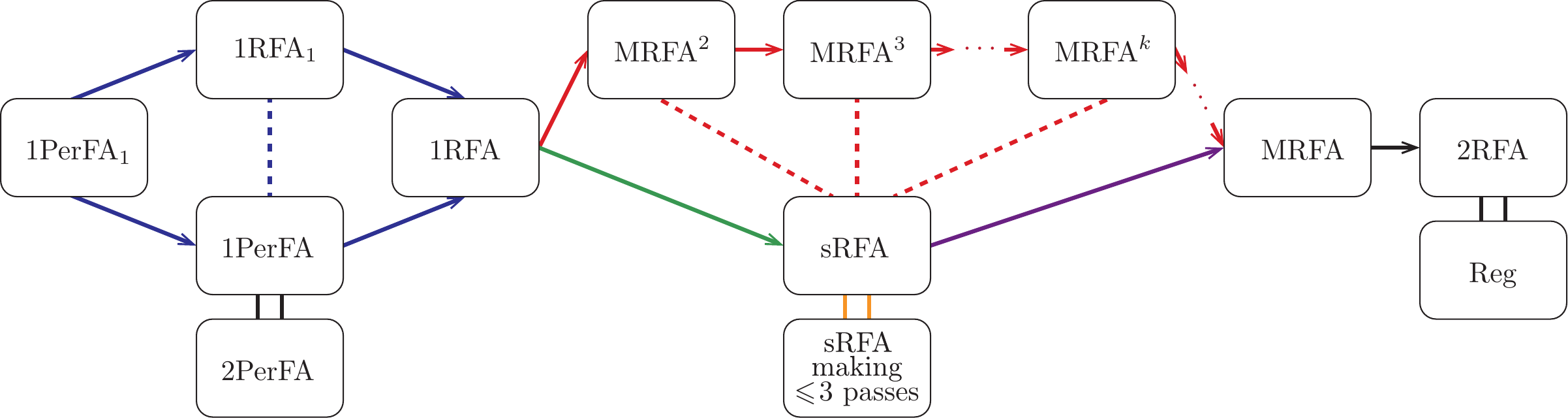}
        \put(101, 165) {\color{mydarkblue} \scriptsize Thm.~\ref{th:one-way-hierarchy}}
        \put(375, 118) {\color{mygreen} \scriptsize Thm.~\ref{th:sRFA-more-expressive-than-1RFA}}
        \put(480, 181) {\color{myred} \scriptsize Thm.~\ref{th:k-entry-MRFA-and-sRFA-comparison}}
        \put(594, 66) {\color{myorange} \scriptsize Thm.~\ref{th:sRFA-restricted-by-the-number-of-passes-no-hierarchy}}
        \put(680, 118) {\color{myviolet} \scriptsize Thm.~\ref{sRFA_subset_MRFA_theorem}}
    \end{overpic}
    }
    \caption{The hierarchy of reversible and permutation automata.}
    \label{f:hierarchy}
\end{figure}
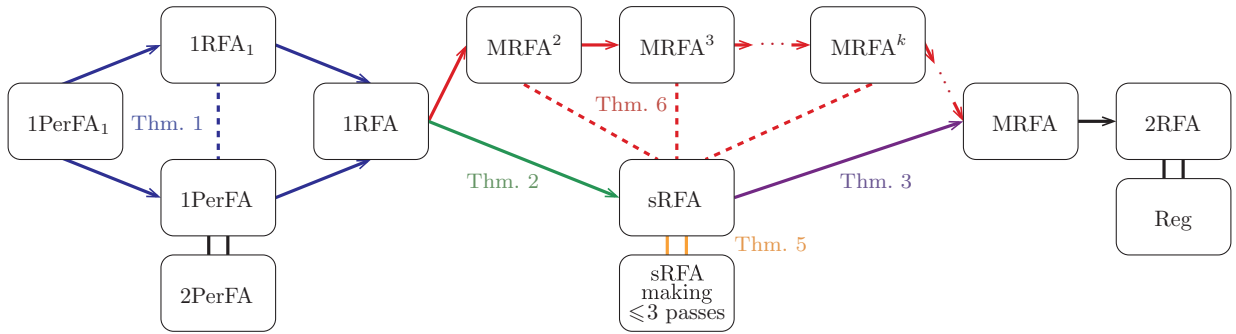

In the paper, we compare the expressive power of these variants of reversible and permutation automata and build a hierarchy. The main results are described in Section~\ref{section:main-hierarchy}. The first result is expected and easy to prove: it is established that reversible automata with one initial state (1RFA$_1$), studied by Angluin~\cite{Angluin}, are incomparable with permutation automata (1PerFA), while reversible automata with unbounded number of accepting states (1RFA) are strictly more powerful than both. Next, 1RFA are compared to reversible sweeping automata (sRFA), and sRFA turn out to be more expressive. A question about relative expressive power of MRFA and sRFA is also studied: sweeping reversible automata can be simulated by reversible automata with multiple initial states, whereas a converse simulation is not always possible.

A hierarchy in the unary case is also investigated in Section~\ref{section:main-hierarchy}. For the majority of models, the results about their expressive power are the same as for unrestricted alphabet. But there are also some differences: it is shown that MRFA over a one-symbol alphabet recognize exactly the same languages as unary sRFA.

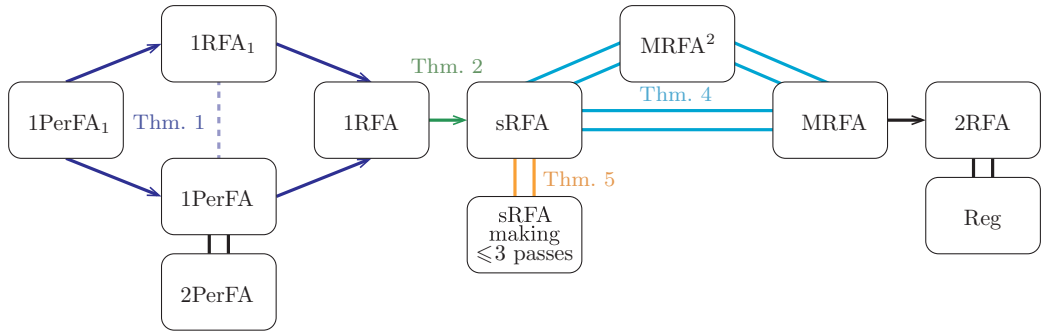
\begin{figure}
    \centerline{%
    \begin{overpic}[scale=.4,permil]{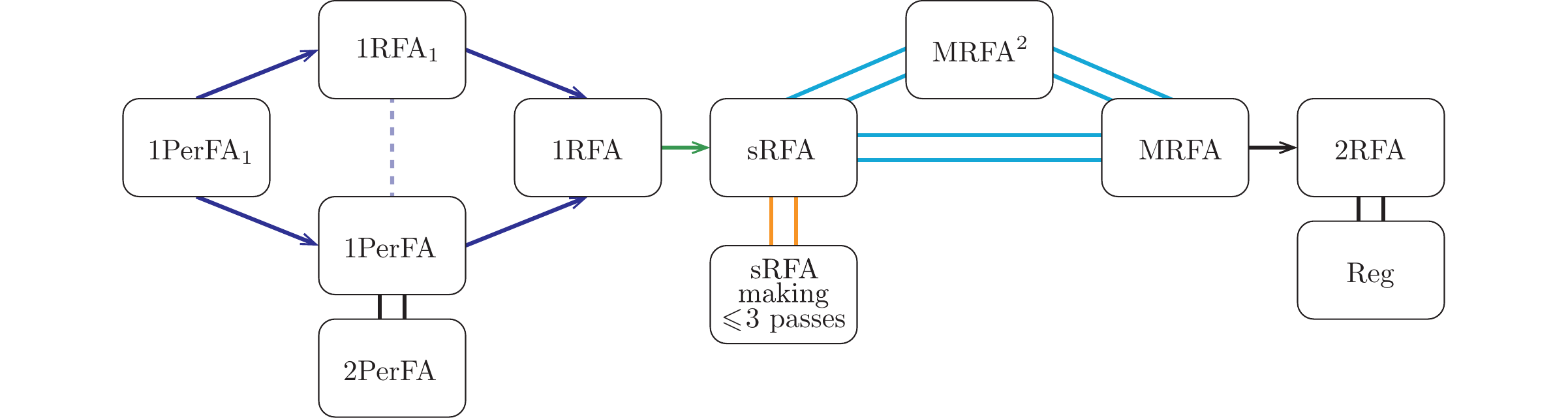}
    \put(180, 165){\color{mydarkblue} \scriptsize Thm.~\ref{th:one-way-hierarchy}}
    \put(407, 211){\color{mygreen} \scriptsize Thm.~\ref{th:sRFA-more-expressive-than-1RFA}}
    \put(515, 118){\color{myorange} \scriptsize Thm.~\ref{th:sRFA-restricted-by-the-number-of-passes-no-hierarchy}}
    \put(594, 188){\color{myblue} \scriptsize Thm.~\ref{th:unary-sRFA-is-the-same-as-unary-MRFA}}
    \end{overpic}
    }%
    \caption{Hierarchy in the unary case.}
    \label{fig:hierarchy-unary}
\end{figure}

It is possible to restrict sweeping automata to make a bounded number of passes over an input string, and this raises a natural question about the existence of a hierarchy by the number of passes, which is addressed in Section~\ref{section:sRFA-restricted-by-the-number-of-passes}. It turns out that this hierarchy collapses, and every sRFA can be transformed to an sRFA making no more than three passes over an input string.

Another succinctness measure worth being investigated is the number of initial states in MRFA. Do MRFA with at most $k$ initial states (MRFA$^k$) form a hierarchy by $k$? In Section~\ref{section:MRFA-hierarchy} it is proved that they actually do, as long as the alphabet contains at least two symbols; furthermore, MRFA$^k$ are incomparable to sRFA for each $k$. In the unary case, this hierarchy collapses.

The resulting hierarchy is shown in Fig.~\ref{f:hierarchy}, and the unary case is illustrated in Fig.~\ref{fig:hierarchy-unary}.

\section{Definitions}\label{section:definitions}

In the literature, different variants of one-way reversible automata were investigated: Angluin~\cite{Angluin} studied reversible automata with one initial and one accepting state (1RFA$_1$), Pin~\cite{Pin1987} allowed reversible automata to have an arbitrary number of initial and accepting states (MRFA), and Holzer et al.~\cite{HolzerJakobiKutrib} studied reversible automata with one initial and any number of accepting states (1RFA). As long as the initial state is unique, reversible automata are a special case of 1DFA.

\begin{definition}
    A one-way deterministic automaton is a quintuple $(\Sigma, Q, q_0, \langle \delta_a \rangle_a, F)$, which includes
    \begin{itemize}
        \item an input alphabet $\Sigma$,
        \item a set of states $Q$,
        \item an initial state $q_0 \in Q$,
        \item a transition function $\delta_a \colon Q \to Q$ by each symbol $a$ from the alphabet $\Sigma$,
        \item a set of accepting states $F \subseteq Q$.
    \end{itemize}
    
    A computation of a 1DFA on a string $s = a_1a_2 \ldots a_k$ is a sequence of states $q_0$, $q_1$, \ldots, $q_k$, in which every next state is obtained by applying the transition function to the previous one: $q_{i + 1} = \delta_{a_{i + 1}}(q_i)$. A one-way deterministic automaton (1DFA) accepts a string if its computation ends in an accepting state $q_k \in F$.
\end{definition}

In the literature, partial deterministic finite automata, with partial transition functions, are also common. Their computation on a string can end in advance by an undefined transition, and in this case the string is rejected.

\begin{definition}
    A one-way reversible automaton (1RFA) is a partial one-way deterministic automaton with an injective transition function $\delta_a$ for each symbol $a$.
\end{definition}

\begin{definition}
    A one-way permutation automaton (1PerFA) is a one-way reversible automaton (1RFA) with a bijective (and hence fully defined) transition function $\delta_a$ by each symbol $a$.
\end{definition}

Sweeping automata, unlike their one-way counterparts, can read a string several times, from left to right and from right to left.

\begin{definition}
    A sweeping deterministic automaton (sDFA) is a 9-tuple $(\Sigma, Q_+, Q_-, q_0, \langle \delta_a^+ \rangle_a, \langle \delta_a^- \rangle_a, \delta_{\vdash}, \delta_{\dashv}, F)$, in which there are
    \begin{itemize}
        \item an input alphabet $\Sigma$,
        \item a set of states $Q_+ \cup Q_-$ split into two disjoint sets, in one of which the automaton reads the input from left to right, and in the other from right to left,
        \item an initial state $q_0 \in Q_+$, in which the automaton starts to read the input,
        \item partial transition functions by each symbol $a \in \Sigma$ inside a string: $\delta_a^+ \colon Q_+ \to Q_+$ for reading from left to right, and $\delta_a^- \colon Q_- \to Q_-$ for reading from right to left,
        \item partial transition functions at the end-markers, $\delta_{\vdash} \colon \{ q_0 \} \cup Q_- \to Q_+$ and $\delta_{\dashv} \colon Q_+ \to Q_-$,
        \item a set of accepting states $F \subseteq Q_+$ effective at the right end-marker.
    \end{itemize}

    A computation of a sweeping deterministic automaton on the string $s = \mathop{\vdash} a_1 a_2 \ldots a_k \mathop{\dashv}$ is a sequence of pairs $(q_j, i_j)$, where $q_j$ is a state, and $i_j$, with $i_j \in \{ 0, 1, \ldots, k, k + 1 \}$, is a position in the string. The elements of the sequence are defined in the following way: the first pair is $(q_0, 0)$, and the $j + 1$-th one is obtained from the $j$-th pair by changing the state and the position of the current symbol according to the transition function.

    If a sweeping automaton's head is inside the string, then it moves the head to the next or to the previous symbol, depending on whether the automaton is in a state from $Q_+$ or from $Q_-$. Also, in this case the automaton uses functions $\delta_{a_i}^+$ and $\delta_{a_i}^-$ to change its state. And if the automaton's head is at the left end-marker, then it moves the head to the first symbol $a_1$ of the string and changes its state according to the transition function $\delta_{\vdash}$. And if the head is at the right end-marker, then the automaton proceeds to the last symbol $a_k$ of the string and changes the state in accordance with $\delta_{\dashv}$.
    \begin{equation*}
        (q_{j + 1}, i_{j + 1}) =
        \begin{cases}
            (\delta_{a_{i_j}}^+(q_j), i_j + 1), &\text{if } 1 \leqslant i_j \leqslant k \text{ and } q_j \in Q_+;
            \\
            (\delta_{a_{i_j}}^-(q_j), i_j - 1), &\text{if } 1 \leqslant i_j \leqslant k \text{ and } q_j \in Q_-;
            \\
            (\delta_{\vdash}(q_j), 1), &\text{if } i_j = 0 \text{ and } q_j \in Q_- \cup \{ q_0 \};
            \\
            (\delta_{\dashv}(q_j), k), &\text{if } i_j = k + 1 \text{ and } q_j \in Q_+.
        \end{cases}
    \end{equation*}

    A sweeping deterministic automaton accepts the string, if the computation ends at the right end-marker in an accepting state, that is, in a pair $(q, k + 1)$ with $q \in F$.
\end{definition}

As in the case of one-way automata, it is possible to define a reversible sweeping automaton. Such an automaton is an sDFA with the restriction that 
its transition functions are injective.

\begin{definition}
    A sweeping reversible automaton (sRFA) is a sweeping deterministic automaton, with injective transition functions inside the string, $\delta_a^+$ and $\delta_a^-$, for each symbol $a \in \Sigma$, and with transition functions by the end-markers, $\delta_{\vdash}$ and $\delta_{\dashv}$, injective as well.
\end{definition}

Furthermore, a sweeping reversible automaton (sRFA) with acceptance at both sides can be defined, which is a more general model of sRFA that can accept at either end-marker. That is, the set of accepting states of an sRFA with acceptance at both sides is $F \subseteq Q$. And the computation of such an automaton on a string $s = \mathop{\vdash} a_1 a_2 \ldots a_k \mathop{\dashv}$ is defined exactly as for ordinary sRFA, and it accepts the string if the computation ends either at the right or left end-marker in an accepting state, that is, in a pair $(q, i)$ with $q \in F$ and $i \in \{ 0, k + 1 \}$.

Such automata turn out to be no more powerful than ordinary sRFA with acceptance at one side.

\begin{lemma}\label{lem:sRFA-both-markers-accepting-equals-to-sRFA}
    For every sRFA with acceptance at both sides, with a set of states $P_+ \cup P_-$, and with accepting states $E \subseteq P_+ \cup P_-$, there exists an ordinary sRFA (with acceptance at one side), which has the set of states $Q_+ \cup Q_-$, where $Q_+ = P_+ \cup \set{p'}{p \in E \cap P_-}$, $Q_- = P_-$, and which recognizes the same language. Here $p'$ stands for a copy of a state $p$.
\end{lemma}
\begin{proof}[A sketch of a proof.]

The desired automaton works exactly as the given one, until it is about to accept at the left end-marker in one of the states $p \in E \cap P_-$. Instead, the new automaton turns at the left end-marker and proceeds to the right end-marker looping in a separate state $p'$, in which it eventually accepts.
\end{proof}

The last model of sweeping automata considered in the paper are \emph{sweeping permutation automata}.

\begin{definition}
    A sweeping permutation automaton (2PerFA) is a sweeping reversible automaton (sRFA) with bijective (and hence fully defined) transition functions inside a string, $\delta_a^+$ and $\delta_a^-$, for each symbol $a \in \Sigma$. Transition functions at both end-markers can still be any injective functions, as in sRFA.
\end{definition}

A few results in the paper are based on transformations of sweeping automata to one-way automata. They are all similar to the classical transformation of sDFA to 1DFA~\cite{Shepherdson}, which works as follows. For a sweeping deterministic automaton (sDFA) with a set of states $Q_- \cup Q_+$, a one-way deterministic automaton (1DFA) is constructed, which has pairs $(p, f)$ for its states, with $p \in Q_+$, and a partial function $f$ mapping $Q_-$ to $Q_+$. The transitions of the 1DFA are defined so that the reachability of a state $(p, f)$ by some string $s$ is equivalent to the next two conditions.
\begin{enumerate}
    \item After reading the string $s$ from left to right, the given sweeping automaton comes to the state $p$.
    \item If the given sDFA starts to read the string $s$ at its last symbol in a state $q \in Q_-$, reads it from right to left, makes a transition by the left end-marker, and next reads $s$ from left to right, then it finishes in the state $f(q) \in Q_+$ (and the value $f(q)$ must be defined in this case). If the computation of the sDFA on the string $s$ ends prematurely by an undefined transition before the automaton reads $s$ twice, as described above, then $f(q)$ is undefined.
\end{enumerate}

Finally, the last type of automata considered in the paper are reversible automata with multiple initial states.
\begin{definition}[Pin~\cite{Pin1987}]
    A one-way reversible automaton with multiple initial states (MRFA) is a quintuple $(\Sigma, Q, Q_0, \langle \delta_a \rangle_a, F)$, in which
    \begin{itemize}
        \item $\Sigma$ is an input alphabet;
        \item $Q$ is a finite set of states;
        \item $Q_0 \subseteq Q$ is a set of initial states;
        \item $\delta_a \colon Q \to Q$, for each symbol $a \in \Sigma$, is an injective partially defined transition function by $a$;
        \item $F \subseteq Q$ is a set of accepting states.
    \end{itemize}

    The computation of an MRFA on a string $s = a_1 a_2 \ldots a_k$ starting in a state $q_0 \in Q_0$ is a sequence of states $q_0$, $q_1$, \ldots, $q_k$, which begins with $q_0$, and in which every next state is obtained from the previous one by using the transition function: $q_i = \delta_{a_i}(q_{i - 1})$ for $1 \leqslant i \leqslant k$. The computation is accepting, if it ends in an accepting state from the set $F$. A string is accepted by an MRFA, if there exists an accepting computation on that string starting in some initial state.
\end{definition}

This is the most general variant of one-way reversible automata, and the following lemma for showing non-representability of languages by these automata is known.

\begin{lemma}[Pin~\cite{Pin1987}]\label{pin_lemma}
Let a language $L$ be recognized by some MRFA, and let $xy^+z \subseteq L$ for some strings $x, z \in \Sigma^*$ and $y \in \Sigma^+$. Then $xz \in L$.
\end{lemma}

For example, for the language $\{ a, b \}^* a$, the lemma is applied to $x = \epsilon$, $y = a$ and $z = \epsilon$, and it provides a contradiction. Hence, this language is not recognized by any MRFA (as well as by 1RFA and by weaker models).

One of the results of this paper will imply that Lemma~\ref{pin_lemma} also applies to all languages recognized by sRFA.

\section{Main hierarchy}\label{section:main-hierarchy}

Let us begin to build the hierarchy with results for the weaker models.

\begin{theorem}\label{th:one-way-hierarchy}
    One-way permutation automata (1PerFA), one-way reversible automata (1RFA), and their variants with one accepting state (1PerFA$_1$, 1RFA$_1$) form the following proper inclusions.
    \begin{align*}
        &\mathcal{L}(\mathrm{1PerFA}_1) \subsetneq \mathcal{L}(\mathrm{1PerFA}) \subsetneq \mathcal{L}(\mathrm{1RFA}),
        \\
        &\mathcal{L}(\mathrm{1PerFA}_1) \subsetneq \mathcal{L}(\mathrm{1RFA}_1) \subsetneq \mathcal{L}(\mathrm{1RFA}),
    \end{align*}
    where $\mathcal{L}(\mathcal{X})$ stands for the family of languages recognized by a class of automata $\mathcal{X}$.
    
    Also, 1PerFA and 1RFA$_1$ are incomparable in the expressive power.
\end{theorem}
\begin{proof}
Reversible automata with one or several accepting states can recognize all finite languages, for example, the language $\{ a \}$. On the other hand, every permutation automaton recognizes an infinite language (or the empty language). Therefore, the language $\{ a \}$ separates the classes of automata, 1PerFA and 1RFA, and also 1PerFA$_1$ and 1RFA$_1$, by the expressive power, and hence, the inclusion between them is proper.

To separate automata with one accepting state from those with several accepting states, let us consider the language $(a^3)^* \cup a(a^3)^*$. It is possible to construct a 1PerFA with two accepting states recognizing this language, which is also a 1RFA. In this automaton, there are three states corresponding to residues modulo three, and the transition function by $a$ forms a cycle on them. States corresponding to residues 0 and 1 are accepting. This language cannot be recognized by permutation nor reversible automata with a single accepting state. Indeed, to recognize this language, a reversible automaton must have a cycle by the symbol $a$ containing the initial state, because there are arbitrarily long strings in the language. This cannot be a cycle of length one, because then the language recognized by the automaton would either be empty or $a^*$. In this cycle, after reading strings $aaa$ and $aaaa$ from the language, the automaton ends in two neighboring states, which must both be accepting. Hence, every 1RFA (and, in particular, every 1PerFA) recognizing $(a^3)^* \cup a(a^3)^*$ must have at least two accepting states. That is, this language indeed separates the classes 1PerFA$_1$ and 1PerFA, and also 1RFA$_1$ and 1RFA, by the expressive power.

It remains to show, why 1PerFA and 1RFA$_1$ are incomparable in the expressive power. It was proved above, that the language $(a^3)^* \cup a(a^3)^*$ is recognized by 1PerFA, but is not recognized by 1RFA$_1$. Also, it is true, that 1RFA$_1$ can recognize finite languages, for example, the language $\{ a \}$, while 1PerFA can recognize only infinite languages. Hence, the languages $(a^3)^* \cup a(a^3)^*$ and $\{ a \}$ separate 1PerFA and 1RFA$_1$ by the expressive power, and thus incomparability of these classes is obtained.
\end{proof}

Recently, it was shown that sweeping permutation automata (2PerFA) are equal in their expressive power to their one-way variant (1PerFA)~\cite{RadionovaOkhotin}. Could it also be possible to transform every sRFA to a 1RFA? The answer turns out to be negative. What is the difficulty of transforming sRFA to 1RFA? That is, why does the usual transformation of sRFA to 1DFA, in which the constructed one-way automaton computes a certain behavior function that encodes computations of the sweeping automaton on the prefix read so far, sometimes produce an irreversible automaton? This is because a reversible sweeping automaton can reject in the middle of a string, thus invalidating some earlier computed values of the behavior function. Therefore, the same behavior function can be obtained from two different behavior functions upon reading a symbol with undefined transitions. The following example shows that a transformation of sRFA to 1RFA is impossible in the general case.

\begin{example}\label{even_and_a_1rfa_but_not_srfa_example}
    The language $(aa)^* \cup \{ a \}$ cannot be recognized by any one-way reversible automaton (1RFA), but it is recognized by a sweeping reversible automaton (sRFA).
\end{example}
\begin{proof}
Indeed, to recognize the language, a 1RFA must have a cycle by the symbol $a$, which contains the initial state. The initial state is accepting, because the automaton accepts the empty string. Hence, the length of the cycle is even. But, at the same time, the state $\delta_a(q_0)$ is accepting too, as the string $a$ is in the language. Thus, the automaton also accepts a long string with an odd number of symbols, a contradiction.

\begin{figure}[t]
    \centerline{\includegraphics[scale=0.4]{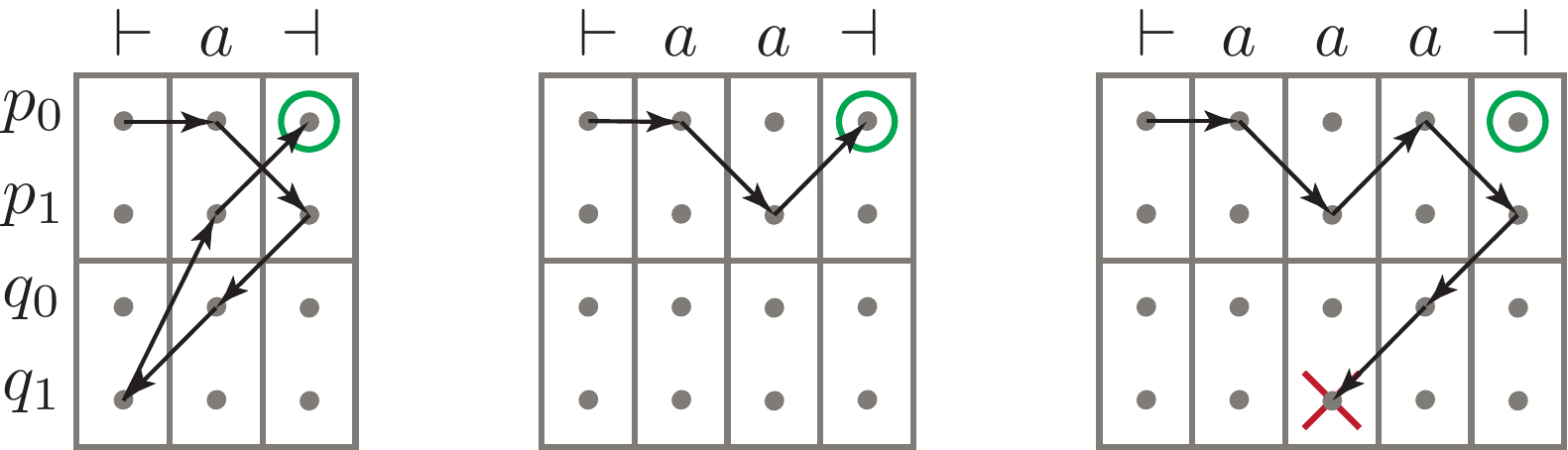}}
    \caption{%
    Computations of an sRFA recognizing the language $(aa)^* \cup \{ a \}$.}
    \label{f:srfa_but_not_1rfa}
\end{figure}

Now, let us build an sRFA recognizing this language. In $Q_+$, it has two states, $p_0$ and $p_1$, and transitions between them form a cycle by $a$. The initial state is accepting, and the automaton recognizes all even strings in it. Only the string $a$ remains, and to accept it, the automaton turns at the right end-marker from the state $\delta_a^+(p_0)$ to the state $q_0 \in Q_-$. In $Q_-$, the automaton also has two states, and a single transition by $a$ is directed from $q_0$ to $q_1$. The sweeping automaton reads one symbol, coming to the state $q_1$, and if the string ends, then the automaton turns at the left end-marker and comes to the state $p_1$, as shown in Fig.~\ref{f:srfa_but_not_1rfa} (at the left), and finally it comes to the accepting state at the next step, and that is how it accepts $a$.
The automaton cannot accept anything extra, because after the first pass it can immediately accept only strings of even length, and if it goes to the second pass, then it can accept the string $a$ only.
\end{proof}

From the example above, the next theorem is obtained.

\begin{theorem}\label{th:sRFA-more-expressive-than-1RFA}
    One-way reversible automata (1RFA) recognize fewer languages than sweeping reversible automata (sRFA).
\end{theorem}

Turning back to the transformation of sRFA to one-way automata, why does the classic transformation create irreversible transitions for the sRFA from Example~\ref{even_and_a_1rfa_but_not_srfa_example}? When a one-way deterministic automaton (1DFA) starts to read a string, it does not know in advance whether the sRFA will return to the left end-marker in its computation and make a turn there. The possibility of returning cannot be excluded, hence, the 1DFA should remember this transition. But after reading the second symbol $a$ it becomes clear, that such a transition is not in the computation, thus, the 1DFA safely forgets it (thereby it remembers, that the string contains more than one symbol). Then, after two steps, it returns to the same state with the forgotten sRFA's transition at the left end-marker, and this is an irreversible transition of the 1DFA.

How to avoid such problems with irreversibility? To relieve a one-way automaton from the burden of forgetting undefined transitions of sRFA, let it guess non-deterministically the domain of a behavior function before starting a computation, and then watch that its size never decreases. If the domain is ever reduced upon reading a symbol, then the automaton just rejects at that point.

This observation allows an sRFA to be transformed to a one-way reversible automaton with multiple initial states (MRFA), which guesses a behavior function's domain at the beginning of its computation, and then proceeds with reversibly calculating the function, maintaining the size of its domain.

\begin{lemma}\label{lem:sRFA-to-MRFA}
    A sweeping reversible automaton (sRFA) with sets of states $P_+$ and $P_-$ can be transformed to a one-way reversible automaton with multiple initial states (MRFA) with the set of states
    \begin{equation*}
        Q = \setbig{(p, f)}{p \in P_+, \: f \colon P_- \to P_+, \: p \notin \Image f}
    \end{equation*}
\end{lemma}
\begin{proof}
    Given a sweeping reversible automaton $\mathcal{A} = (\Sigma, P_+, P_-, p_0, \langle \gamma_a^+ \rangle_a, \langle \gamma_a^- \rangle_a, \gamma_{\vdash}, \gamma_{\dashv}, E)$, a one-way reversible automaton with multiple initial states $\mathcal{B} = (\Sigma, Q, Q_0, \langle \delta_a \rangle_a, F)$ is constructed, which guesses the domain of the behavior function of the sweeping automaton on a string while choosing the initial state. Then the MRFA calculates this function while reading the string, and in the end determines the result of the sweeping automaton's computation using that function, and accordingly makes a decision to accept the string or not. The automaton manages to be reversible through using multiple initial states. In each computation, it now has no need to reduce the domain of the constructed behavior function: if such a moment comes, then the automaton can just reject by an undefined transition.

    In each of its states, the MRFA stores a state of the given sRFA and its behavior function. Such a pair has the following meaning: if the MRFA comes to a state $(p, f)$ by some string $s$, then the given sRFA comes to the state $p \in P_+$ after reading $s$ for the first time from left to right. And the partial function $f \colon P_- \to P_+$ maps every state $r$ to another state $r'$, such that the sRFA, having started at the last symbol of $s$ in the state $r$, reads $s$ from right to left, and then from left to right after turning at the left end-marker, finishing in the state $r' = f(r)$. Note that, for some string $s$, there may be multiple states $(p, f)$ of the MRFA satisfying the above condition: indeed, the sRFA may potentially make more computations on $s$ than are reflected in $f$. The main idea of the construction is that the MRFA can reach \emph{all} such states, if it starts from different initial states.
    
    \begin{align*}
        Q &= \setbig{(p, f)}{p \in P_+, \: f \colon P_- \to P_+, \: p \notin \Image f}
    \intertext{%
    The MRFA's initial states must satisfy the above property for the empty string $s = \varepsilon$. The first component of every initial state is always set to the initial state $p_0$ of the sRFA, and the second component, a function, is always defined to act as the transition function at the left end-marker $\gamma_{\vdash}$. Moreover, the MRFA chooses the domain of a behavior function of the sRFA in the initial states, hence, for their second component, they have all possible restrictions of the function $\gamma_{\vdash}$ to subsets of its domain.
    }
        Q_0 &= \setbig{(p_0, f)}{f = \gamma_{\vdash}|_S \text{ for some } S \subseteq (\Dom \gamma_{\vdash}) \setminus \{ p_0 \}}
    \intertext{%
    While making a transition by a symbol $a$, the MRFA must change both components of its state. And it changes them so that the new state of the MRFA again describes the behavior of the given sRFA, but on the string extended with the symbol $a$. For this, the transition functions of the sRFA are applied to the pair $(p, f)$, from which the MRFA makes its transition, as shown below. Moreover, if the domain of the function $f$ is reduced after applying the symbol, then the transition is left undefined in order to avoid irreversible behavior.
    }
        \delta_a(p, f) &=
        \begin{cases}
            \big(\gamma_a^+(p), \gamma_a^+ \circ f \circ \gamma_a^-\big), & \text{if } \gamma_a^+(p) \text{ is defined,}
            \\
            & \text{and } |\Dom\gamma_a^+ \circ f \circ \gamma_a^-| = |\Dom f|; \\
            \text{is undefined}, & \text{otherwise}.
        \end{cases}
    \intertext{%
    In the definition of the MRFA's set of accepting states, for each state $(p, f)$, consider the sequence of the sRFA's states, in which it visits the right end-marker in the computation described by the function $f$. The first state from the sequence, $p_1$, is equal to $p$. To compute the second one, apply the transition function at the right end-marker $\gamma_{\dashv}$, and then the function $f$ to the state $p_1$ obtaining $p_2 = f(\gamma_{\dashv}(p_1))$, etc. If the resulting sequence $p_1$, $p_2$, \ldots\ ends in an accepting state of the sRFA, then $(p, f)$ will be accepting in the MRFA.
    }
        F &= \setbig{(p, f)}{(f \circ \gamma_{\dashv})^k(p) \in E \text{ for some } k \geqslant 0}.
    \end{align*}

    Let us prove that the transitions defined above are reversible, that is, the constructed automaton is indeed an MRFA.

    \begin{claim}\label{claim:sRFA_to_MRFA_reversibility}
        If a transition $\delta_a\big((p, f)\big) = (q, g)$ is defined, then $p = (\gamma_a^+)^{-1}(q)$, $f = (\gamma_a^+)^{-1} \circ g \circ (\gamma_a^-)^{-1}$ and $|\Dom f|= |\Dom g|$. In particular, the constructed automaton is reversible.
    \end{claim}
    \begin{proof}
        Let $\delta_a\big((p, f)\big) = (q, g)$. Then, first, $q = \gamma_a^+(p)$, which implies $p = \big(\gamma_a^+\big)^{-1}(q)$, as the function $\gamma_a^+$ is injective. Secondly, $g = \gamma_a^+ \circ f \circ \gamma_a^-$, and the equality $|\Dom g| = |\Dom f|$ holds, that is, the domain of the function $f$ is not reduced after taking a composition with the partial functions $\gamma_a^+$ and $\gamma_a^-$. Now consider the function $g' = \big(\gamma_a^+\big)^{-1} \circ g \circ \big(\gamma_a^-\big)^{-1}$, and our goal is to prove, that it coincides with the function $f$. Let us express $g'$ in the following way:
        \begin{multline*}
            g' = \big(\gamma_a^+\big)^{-1} \circ g \circ \big(\gamma_a^-\big)^{-1} = \big(\gamma_a^+\big)^{-1} \circ \gamma_a^+ \circ f \circ \gamma_a^- \circ \big(\gamma_a^-\big)^{-1}
                = \\ =
            \big( \big(\gamma_a^+\big)^{-1} \circ \gamma_a^+ \big) \circ f \circ \big( \gamma_a^- \circ \big(\gamma_a^-\big)^{-1} \big)
        \end{multline*}
        The transition functions $\gamma_a^+$ and $\gamma_a^-$ are injective, hence, the compositions $\big(\gamma_a^+\big)^{-1} \circ \gamma_a^+$ and $\gamma_a^- \circ \big(\gamma_a^-\big)^{-1}$ are identities on their respective domains. Let us find out, how the domains and ranges of these functions are arranged in comparison with the function $f$. As it is true that $|\Dom \gamma_a^+ \circ f \circ \gamma_a^-| = |\Dom f|$, then $|\Dom f \circ \gamma_a^-| = |\Dom f|$. From this, the inclusion $\Dom f \subseteq \Image \gamma_a^-$ is obtained, because otherwise the domain of the composition of these functions has a smaller size than the domain of $f$. From the equality $|\Dom \gamma_a^+ \circ f \circ \gamma_a^-| = |\Dom f \circ \gamma_a^-|$, the inclusion $\Image f \subseteq \Dom \gamma_a^+$ is obtained on the similar grounds. Returning to the compositions $\big(\gamma_a^+\big)^{-1} \circ \gamma_a^+$ and $\gamma_a^- \circ \big(\gamma_a^-\big)^{-1}$, it can be written now, that
        \begin{equation*}
            \Dom f \subseteq \Image \gamma_a^- = \Image \gamma_a^- \circ \big(\gamma_a^-\big)^{-1},
        \end{equation*}
        \begin{equation*}
            \Image f \subseteq \Dom \gamma_a^+ = \Dom \big(\gamma_a^+\big)^{-1} \circ \gamma_a^+.
        \end{equation*}
        It is first obtained that $\Dom f$ is a subset of the image of the function $\gamma_a^- \circ \big(\gamma_a^-\big)^{-1}$, with which the composition at the right is taken in the definition of the function $g'$. And secondly, $\Image f$ is a subset of the domain of the function $\big(\gamma_a^+\big)^{-1} \circ \gamma_a^+$, with which the composition is taken at the left in the definition of $g'$. Moreover, both functions are identities on their domains. Hence, the composition $g' = \big( \big(\gamma_a^+\big)^{-1} \circ \gamma_a^+ \big) \circ f \circ \big( \gamma_a^- \circ \big(\gamma_a^-\big)^{-1} \big)$ is indeed equal to the function $f$.
    \end{proof}
    Let us prove that the constructed MRFA indeed recognizes the same language as the given sRFA. For this, the next claim is needed.
    
    \begin{claim}\label{claim:MRFA-s-good-states-reachability}
        On a string $s$, the constructed MRFA may come to a state $(p, f)$
        from any initial state if and only if
        \begin{enumerate}
            \item the sRFA comes to the state $p \in P_+$ after the first pass on $s$ from left to right;
            \item the partial injective function $f \colon P_- \to P_+$ maps each state $q \in P_-$ from its domain to such a state $f(q) \in P_+$ that when the sRFA starts to read $s$ from right to left in the state $q$, then turns at the left end-marker and reads $s$ once again from left to right, it finishes in the state $f(q)$. Also, $p$ is not in the image of $f$.
        \end{enumerate}
    \end{claim}
    \begin{proof}
    \textcircled{$\Rightarrow$}
    First, it is proved by induction on the length of a string that all states reachable by $s$ satisfy both properties. For the empty string, the claim holds: the MRFA starts and finishes reading $\varepsilon$ in one of the states from $Q_0$, and in each of them, the first component is the state $p_0$, which satisfies the first condition, and the second component is a restriction of $\gamma_{\vdash}$ to some subset of $Q_-$, and it satisfies the second property.

    Now let us prove the induction step for the string $v = ua$. Let the MRFA finish reading $v$ in a state $(q, g)$. In its computation, at the penultimate step the automaton has read the whole string $u$ and has reached some state $(p, f)$, which satisfies both properties by the induction hypothesis. Next, it reads $a$ and comes to the state $(q, g)$, which is defined as
    \begin{equation*}
        (q, g) = (\gamma_a^+(p), \gamma_a^+ \circ f \circ \gamma_a^-).
    \end{equation*}
    Let us prove both conditions for the state $(q, g)$. By the induction hypothesis, it is true that the sRFA finished the first reading of $u$ from left to right in the state $p$, and from this it follows that the first reading of $v = ua$ from left to right finishes in $\gamma_a^+(p)$. To prove the second property for the function $\gamma_a^+ \circ f \circ \gamma_a^-$, notice that a state $r \in P_-$ from its domain is mapped to $\gamma_a^-(r) \in P_-$ first, and this corresponds to the sRFA's reading the last symbol of $v = ua$ from right to left. Then, the function $f$ is applied, and it maps a state $\gamma_a^-(r)$ to the state $f(\gamma_a^-(r)) \in P_+$, in which the sRFA comes after reading the prefix $u$ of the string $v$ from right to left starting in $\gamma_a^-(r)$, and then from left to right. And next, the transition function by $a$ is applied, using which the sRFA finishes reading $v = ua$ from left to right, and the equality $g(r) = \gamma_a^+(f(\gamma_a^-(r)))$ is obtained.
    
    The function $g$ is injective, because such is the function $f$ and the transition functions $\gamma_a^+$, $\gamma_a^-$. The function $g$ does not contain $q$ in its image, because $\gamma_a^+$, which we applied to both the function $f$ and the state $p$ to obtain the current pair $(q, g)$, is injective and maps all the values from the image of $f$ and the state $p$ to pairwise distinct states.
    
    \textcircled{$\Leftarrow$}
    Now, let us prove, that if a string $s$ is fixed, then every state $(p, f)$ satisfying both properties from the theorem is reachable in the MRFA after reading $s$ from some initial state. The proof is again by induction on the length of a string. For the empty string, the first condition means $p = p_0$, and the second condition is equivalent to the function $f$ acting on its domain as $\gamma_{\vdash}$. All such states $(p, f)$ are initial states of the constructed MRFA, in which it can come after reading the empty string.

    Let us prove the induction step for the string $v = ua$. Let a state $(q, g)$ satisfy both conditions for the string $ua$. The first property says, that the sRFA finishes the first reading of the string $v$ from left to right in the state $q$. As this automaton is reversible, it is possible to restore the state, in which the sRFA was on the previous step, and it is $p = (\gamma_a^+)^{-1}(q)$. In this state, the sRFA finishes the first reading of $u$ from left to right.

    Now, consider the function $g$ satisfying the second property. Notice that from each state $r \in P_-$, on which the function $g$ is defined, the given sRFA first comes to the state $\gamma_a^-(r)$ in the computation on the string $ua$. Then, after several steps of the computation corresponding to reading the prefix $u$ first from right to left, and then from left to right, the sRFA comes to some state $r' \in P_+$. Next, in the computation on $v = ua$ the last transition by the symbol $a$ is applied, the state $r'$ is changed to $\gamma_a^+(r')$, which coincides with the value of the function $g(r) = \gamma_a^+(r')$. Now it is possible to define a partial function $f \colon P_- \to P_+$, as shown in Fig.~\ref{fig:restoring-f-from-g}, which maps $\gamma_a^-(r)$ to $r'$ for every state $r \in P_- \cap \Dom g$, as described above. The domain of $f$ is exactly the set $\set{\gamma_a^-(r)}{r \in \Dom g}$. It is true that $|\Dom f| = |\Dom g|$, that is, the size of the domain is the same as the size of the domain of $g$,because there is a bijection between the sets, which is the function $\gamma_a^-$ restricted to the domain of the function $g$.
    And also $g = \gamma_a^+ \circ f \circ \gamma_a^-$ holds by the definition of the function $f$.

    \begin{figure}[h!]
        \centering
        \includegraphics[scale=0.4]{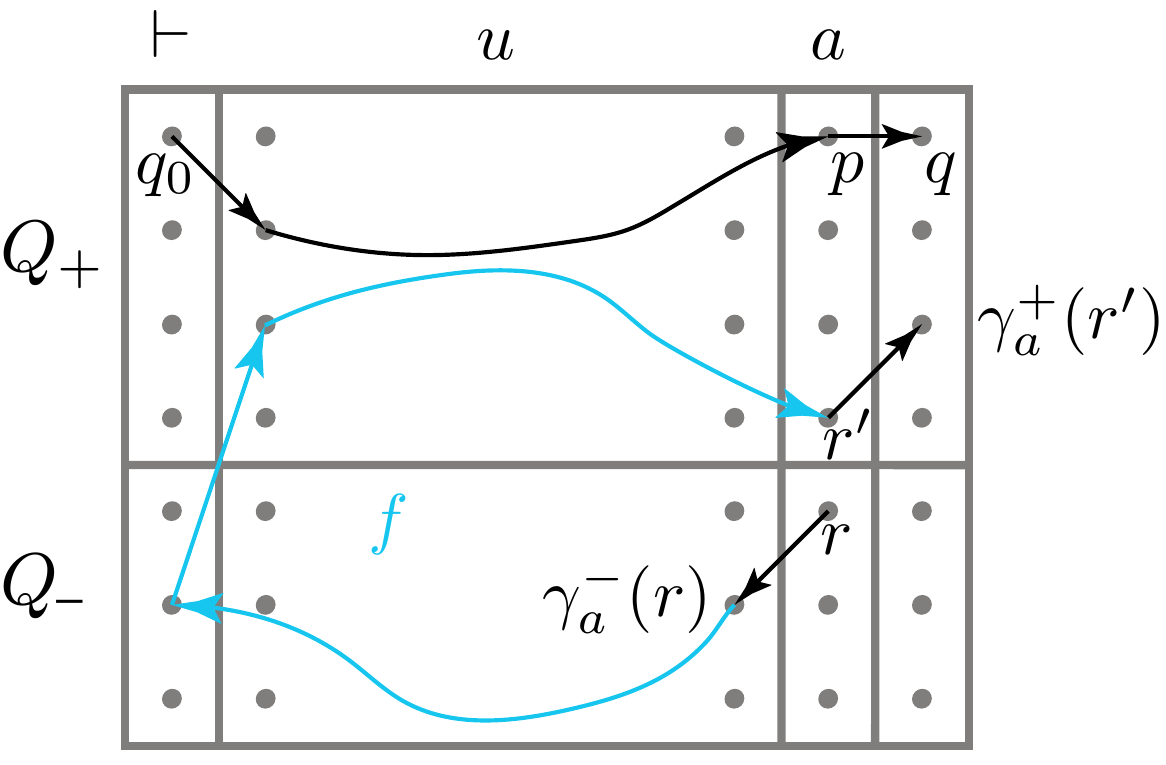}
        \caption{Constructing the function $f$ from the function $g$.}
        \label{fig:restoring-f-from-g}
    \end{figure}

    It is obtained, that the pair $(p, f)$ satisfies both properties for the string $u$, and hence, by the induction hypothesis, this pair is reachable in the MRFA by the string $u$ from one of its initial states. Also, it is known, that $q = \gamma_a^+(p)$, $g = \gamma_a^+ \circ f \circ \gamma_a^-$ and $|\Dom f| = |\Dom g|$, thus, there is a transition from $(p, f)$ to $(q, g)$ by the symbol $a$ in the MRFA by the definition. And from this, it follows that the state $(q, g)$ is indeed reachable by the string $v$.
    \end{proof}
    Now let us show the mutual containment between the languages recognized by the sRFA and by the constructed MRFA. For the given string $s$, the computation of the sRFA on it is accepting if and only if there exists a pair $(p, f)$ satisfying both properties from Claim~\ref{claim:MRFA-s-good-states-reachability} and satisfying the definition of the accepting states of the MRFA. Indeed, such a pair $(p, f)$ encodes a correct accepting computation of the sRFA on the string $s$, and also, every accepting computation of the sRFA on a string can be encoded in such a pair. The existence of the pair $(p, f)$ satisfying both conditions from Claim~\ref{claim:MRFA-s-good-states-reachability} is equivalent to $(p, f)$ being reachable in the MRFA by the string $s$ from some initial state. As the pair $(p, f)$ is also an accepting state of the MRFA, its existence is equivalent to $s$ being accepted by MRFA.
\end{proof}

To show that sRFA are strictly weaker than MRFA, consider the following example.

\begin{figure}
    \centering
    \includegraphics[scale=0.4]{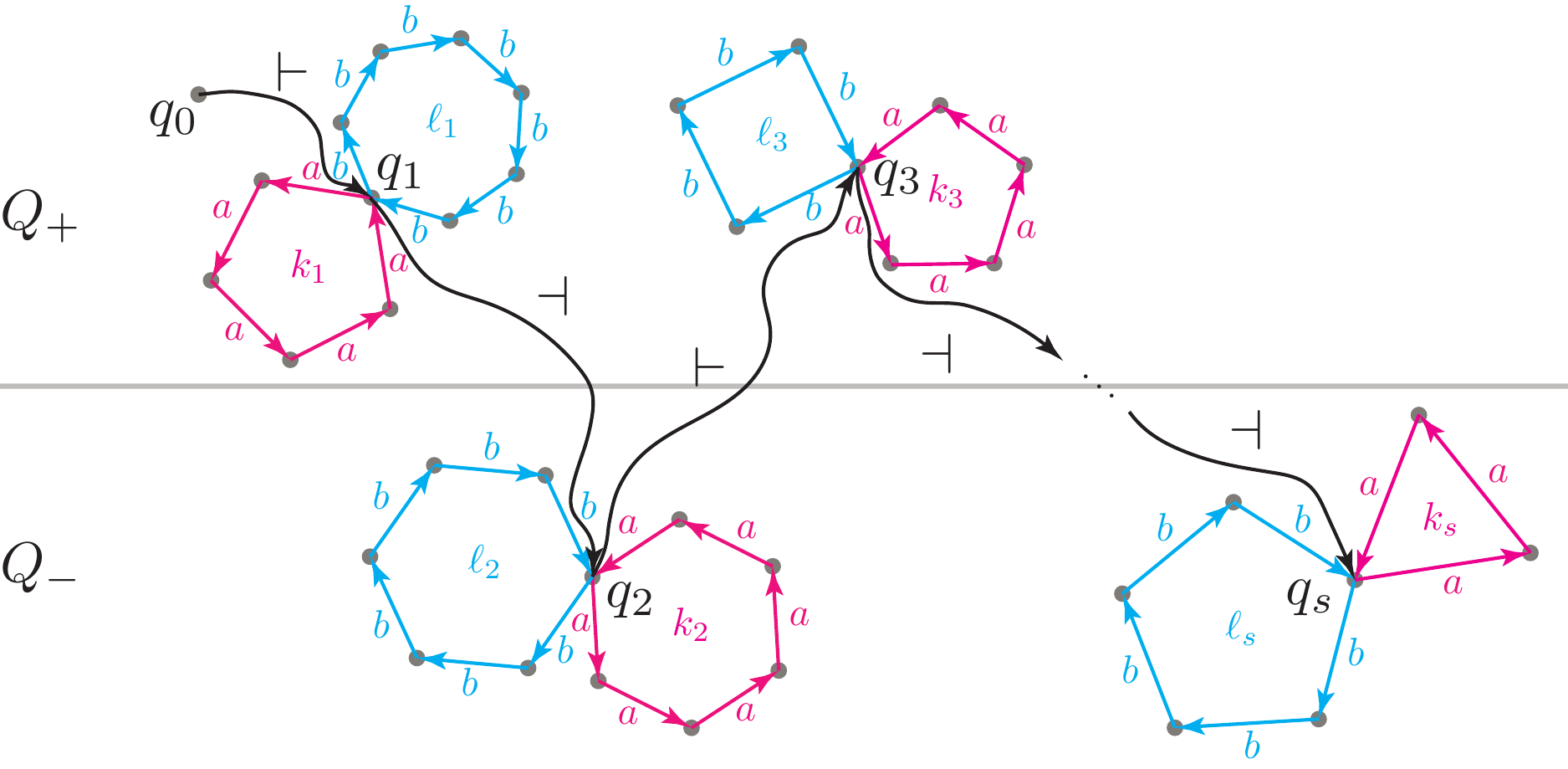}
    \caption{Cycles in sRFA}
    \label{fig:sRFA-does-not-recognize-ab}
\end{figure}
\begin{example}\label{example:a*-cup-b*}
    The language $a^* \cup b^*$ is recognized by an MRFA, but no sRFA can recognize it.
\end{example}
\begin{proof}
    Indeed, suppose that there is a sweeping reversible automaton (sRFA) recognizing this language, and let $Q_+ \cup Q_-$ be its set of states.
    In its initial configuration, it makes a transition at the left end-marker, entering some state $q_1 \in Q_+$. Then, while reading a string consisting of symbols of the same type, the automaton loops in the state $q_1$.
    This is a cycle of length $k_1$ by the symbol $a$, and another cycle of length $\ell_1$ by the symbol $b$. The state $q_1$ cannot be rejecting at the right end-marker, but it also cannot be an accepting one, because otherwise the automaton accepts the string $a^{k_1}b^{\ell_1}$. Hence, the automaton turns at the right end-marker in $q_1$ and comes to a state $q_2 \in Q_-$. The automaton comes to this state after reading every string of the form $a^{nk_1}$ and $b^{n\ell_1}$, for any natural number $n$. On the way back, while reading sufficiently long strings of such form, the automaton loops again in the state $q_2$. These are a cycle by the symbol $a$ of length $k_2$, and a cycle by the symbol $b$ of length $\ell_2$. From the state $q_2$, the automaton again has to turn at the left end-marker and come to some state $q_3 \in Q_+$. And so on, repeating this method of cycle detection, the sequence of states $q_1$, $q_2$, $q_3$, \ldots, in which the automaton is after the turns at the end-markers, is obtained, as shown in Fig.~\ref{fig:sRFA-does-not-recognize-ab}. This sequence is finite, because the automaton is reversible and its computation cannot loop. In every state $q_i$, there are two cycles: the first one is by the symbol $a$ of length $k_i$, and the second one is by the symbol $b$ of length $\ell_i$. Hence, in the last state $q_s$ of the sequence, the automaton must accept or reject, but in this way it either rejects strings consisting only of symbols $a$, which are in the language, or accepts the string $a^mb^n$, where $m = \text{lcm}(k_1, \ldots, k_s)$ and $n = \text{lcm}(\ell_1, \ldots, \ell_s)$, which is not in the language.
\end{proof}

Now, from Lemma~\ref{lem:sRFA-to-MRFA} and Example~\ref{example:a*-cup-b*} follows a theorem about comparing the expressive power of sRFA and MRFA.
\begin{theorem}\label{sRFA_subset_MRFA_theorem}
    Sweeping reversible automata (sRFA) recognize strictly fewer languages than one-way reversible automata with multiple initial states (MRFA).
\end{theorem}
In the unary case, it turns out that sRFA and MRFA are inseparable in the expressive power, that is, they recognize the same family of languages.
\begin{theorem}\label{th:unary-sRFA-is-the-same-as-unary-MRFA}
    Unary sRFA recognize exactly the same languages as unary MRFA.
\end{theorem}
\begin{proof}
In the previous theorem, it was shown that every language recognized by an sRFA can be recognized by some MRFA. For the unary case, let us show the converse inclusion: sRFA can recognize every language over a one-symbol alphabet recognized by MRFA.

First, let us represent a language $L$ recognized by an MRFA $\mathcal{A}$ as a union of languages recognized by reversible automata with one initial state (1RFA): $L(\mathcal{A}) = L(\mathcal{A}_1) \cup \ldots \cup L(\mathcal{A}_k)$. As each $\mathcal{A}_i$ works over a unary alphabet, the transitions by the single symbol in each automaton form either a cycle containing the initial state, or a simple path from the initial state to a state with an undefined transition. Automata with transitions forming a cycle are permutation automata, hence, it is correct to replace their union with a single permutation automaton $\mathcal{B} = (\{ a \}, P, p_0, \langle \gamma_a \rangle_a, E)$. The rest of the automata from the union recognize finite languages, thus, instead of their union, one can write just a finite language $L^{\leqslant \ell}$. In total:
\begin{equation*}
    L(\mathcal{A}) = L(\mathcal{B}) \cup L^{\leqslant \ell}.
\end{equation*}
The language $(aa)^* \cup \{ a \}$ is of the above form,
and the construction of a sweeping reversible automaton (sRFA) for this language given in Example~\ref{even_and_a_1rfa_but_not_srfa_example}
shall now be generalized.

Construct an sRFA with acceptance at both end-markers recognizing $L(\mathcal{A})$. At the first reading of a string $s$ from left to right in states from $Q_+$, the sRFA tests whether the string $s$ is in the language of $\mathcal{B}$. To check this, the sRFA executes the transitions of $\mathcal{B}$ in the states from $Q_+$.
\begin{align*}
    Q_+ &= P
    \\
    q_0 &= p_0
    \\
    \delta_a^+(p) &= \gamma_a(p)
    \\
    F \cap Q_+ &= E
\end{align*}
Next, the sRFA accepts the string at the right end-marker if it is from $L(\mathcal{B})$, and otherwise it continues its computation to test whether the string belongs to the language $L^{\leqslant \ell}$.
The automaton checks this while reading the string from right to left in the states from $Q_-$, and transitions there form a simple path of $\ell + 1$ states. The automaton makes a final decision whether to accept the string or not after returning to the left end-marker.
\begin{align*}
    Q_- &= \{ q_1, q_2, \ldots, q_{\ell + 1} \}.
    \\
    \delta_a^-(q_i) &=
    \begin{cases}
        q_{i + 1}, &\text{if } i \leqslant \ell,
        \\
        \text{is undefined}, &\text{if } i > \ell.
    \end{cases}
    \\
    F \cap Q_- &= \set{q_i}{a^{i - 1} \in L(\mathcal{A})}.
\end{align*}
\end{proof}

\section{Is there a hierarchy of sRFA by the number of passes over an input string?}\label{section:sRFA-restricted-by-the-number-of-passes}

\begin{theorem}\label{th:sRFA-restricted-by-the-number-of-passes-no-hierarchy}
    Sweeping reversible automata (sRFA) recognize exactly the same languages as sweeping reversible automata making at most three passes over a string (and the same as sweeping automata with acceptance at both sides making at most two passes).
\end{theorem}
\begin{proof}
    Let us transform the given sweeping automaton $\mathcal{A} =$ $(\Sigma, P_+, P_-, p_0, \langle \gamma_a^+ \rangle_a, \langle \gamma_a^- \rangle_a, \gamma_{\vdash}, \gamma_{\dashv}, E)$ to another sRFA $\mathcal{B} = (\Sigma, Q_+, Q_-, q_0, \langle \delta_a^+ \rangle_a, \langle \delta_a^- \rangle_a, \delta_{\vdash}, \delta_{\dashv},$ $F)$, which makes at most three passes over its input. For this, consider a sweeping permutation automaton (2PerFA) $\widetilde{\mathcal{A}}$, which is obtained from $\mathcal{A}$ by completing the transition functions $\gamma_a^+$ and $\gamma_a^-$ by each symbol $a$ to bijections $\widetilde{\gamma}_a^+$ and $\widetilde{\gamma}_a^-$ arbitrarily. The automaton $\mathcal{B}$, while reading a string from left to right for the first time, computes the behavior function $f \colon P_- \to P_+$ of the sweeping permutation automaton $\widetilde{\mathcal{A}}$ on the prefix read, and remembers a state $p \in P_+$, in which $\widetilde{\mathcal{A}}$ comes after reading the current prefix of a string from left to right from the initial state $p_0$.

    The pair $(p, f)$ encodes the behavior of the given sRFA on a string $s$ in the same way as done in the states of the MRFA described in Lemma~\ref{lem:sRFA-to-MRFA}. That is, the function $f$ is such that it maps every state $p$ from $P_-$ to such a state $p'$ from $P_+$, that if the automaton $\widetilde{\mathcal{A}}$ starts at the last symbol of $s$ in the state $p$, then reads the whole string once from right to left in states from $P_-$ and then, after a turn, another time from left to right in states from $P_+$, then it finishes in the state $p'$. In the present construction, due to the switch to a 2PerFA, there is no need for a non-deterministic choice of the initial state, which was essential in the transformation of sRFA to MRFA for avoiding irreversible forgetting of values of $f$. Indeed, the behavior function $f$ of the permutation automaton is injective and defined on a fixed number of states, which is the number of states from $P_-$ with a transition at the left end-marker defined. All transitions of the 2PerFA are bijective, and hence none of them affect the domain of $f$.

    While the sRFA $\mathcal{B}$ computes the behavior function of the permutation automaton in the second component, it makes transitions of the given reversible automaton $\mathcal{A}$ in the first component, and if $\mathcal{B}$ meets an undefined transition of $\mathcal{A}$, then it immediately rejects the input.
    \begin{align*}
        Q_+ &\supseteq \set{(p, f)}{p \in P_+, \: f \colon P_- \to P_+ \text{ is injective partial function}},
        \\
        \delta_a^+(p, f) &=
        \begin{cases}
            (\gamma_a^+(p), \widetilde{\gamma}_a^+ \circ f \circ \widetilde{\gamma}_a^-), & \text{if } \gamma_a^+(p) \text{ is defined in } \mathcal{A}; \\
            \text{undefined}, & \text{otherwise}.
        \end{cases}
    \end{align*}
    The following correctness statement for the first pass of the new sRFA holds true.
    \begin{claim}
        The constructed automaton $\mathcal{B}$ reads a string $w$ completely from left to right if and only if the given automaton $\mathcal{A}$ reads it completely from left to right. Furthermore, if $\mathcal{B}$ reads $w$ completely from left to right, then afterwards it comes to a state $(p, f)$ such that
        \begin{enumerate}
            \item the automaton $\mathcal{A}$ comes to the state $p$ after the first pass over $w$ from left to right;
            \item the injective partial function $f$ encodes the computation of the permutation automaton $\widetilde{A}$ on the string $w$. Namely, every state $q$ from $P_-$, on which the function $f$ is defined, is mapped to the state in which the automaton $\widetilde{A}$ ends its computation after reading the string $w$ from right to left starting in the state $q$, turning at the left end-marker and reading the string $w$ once again from left to right. The function $f$ is undefined on exactly those states, from which the computation of the automaton $\widetilde{A}$ on the string $w$ is interrupted immediately after the first pass from left to right by an undefined transition at the left end-marker.
        \end{enumerate}
    \end{claim}

    In the end of the first pass, at the right end-marker, the automaton is in a state $(p, f)$, and it checks using the behavior function $f$ whether a rejecting computation of the automaton $\widetilde{\mathcal{A}}$ is obtained. A computation is simulated in the same way as in the transformation of sRFA to MRFA, see Lemma~\ref{lem:sRFA-to-MRFA}. First, the state $p_1 = p$ is taken, then the transition at the right end-marker is made, next the behavior function $f$ is applied to it, and the state $p_2$ from $P_+$ is obtained. Then again the transition at the right end-marker and the behavior function are applied, and so on, until an accepting or a rejecting state at the right end-marker is reached. The simulated computation does not loop, as $\widetilde{\mathcal{A}}$ is reversible, hence, the computation can be described by the finite sequence of states $p_1, p_2, \ldots, p_s$ at the right end-marker, which are obtained during the simulation. If this sequence ends with a rejecting state, then the permutation automaton $\widetilde{\mathcal{A}}$ rejects, and then the reversible automaton $\mathcal{A}$ definitely rejects as well, and hence the automaton $\mathcal{B}$ rejects with a clear conscience.
    
    Otherwise, if $\widetilde{\mathcal{A}}$ accepts, then $\mathcal{B}$ would not hastily accept, as did the MRFA from the transformation from Lemma~\ref{lem:sRFA-to-MRFA}, because it knows, that any of the following two possibilities may be true. First, the given reversible automaton $\mathcal{A}$ can perform the same sequence of transitions and indeed accept the input string. Secondly, at some point $\mathcal{A}$ may encounter an undefined transition by one of symbols inside a string while computing along this sequence, and reject in accordance. The sweeping automaton $\mathcal{B}$ has to check which of these two possibilities is the case.

    For this, the automaton $\mathcal{B}$ tests on the way back, that in the computation on a string, which is encoded in the function $f$, there are no undefined transitions of $\mathcal{A}$. For this, it maintains a set $R \subseteq P_-$ of states from $P_-$, which the given sweeping automaton $\mathcal{A}$ must visit while reading an input string from the language, if it is to accept. Initially, the subset $R_f = \{ \gamma_{\dashv}(p_1), \gamma_{\dashv}(p_2), \ldots, \gamma_{\dashv}(p_{s - 1}) \}$ is taken, where $p = p_1, p_2, \ldots, p_{s - 1}$ is the sequence of states obtained during the simulation of the computation of the automaton $\widetilde{\mathcal{A}}$ using the function $f$.
    \begin{equation*}
        \delta_{\dashv}\big((p, f)\big) =
        \begin{cases}
            (f, R_f), & \text{if } (f \circ \gamma_{\dashv})^k(p) \in E \text{ for some } k \geqslant 0;
            \\
            \text{is undefined}, & \text{otherwise}.
        \end{cases}
    \end{equation*}
    At the second pass, $\mathcal{B}$ stores in its state a function and a set of states from $P_-$, which must appear in the computation of the given sRFA $\mathcal{A}$ on the current symbol. The automaton $\mathcal{B}$ checks if there are no undefined transitions in this computation anywhere, either in $P_-$ or in $P_+$. If $\mathcal{B}$ finds such a transition, then it immediately rejects the input string. Also, the function $f$ is no more useful for the automaton, thus, $\mathcal{B}$ reversibly forgets it in the course of the second pass.

    For each state $(f, R)$, the automaton $\mathcal{B}$ at first finds states of $\mathcal{A}$ at the next step on its way to the left, trying to make a transition by a symbol $a$.
    
    \begin{align*}
        R' &= \gamma_a^-(R).
    \intertext{%
    Next, from the function $f$, the behavior function of $\widetilde{\mathcal{A}}$ on the prefix without the symbol $a$ is restored.
    }
        f' &= \left( \widetilde{\gamma}_a^+ \right)^{-1} \circ f \circ \left( \widetilde{\gamma}_a^- \right)^{-1}
    \intertext{%
    Finally, the transition is not defined if either of the sets of reachable states on the way to the left and on the way to the right are reduced, that is, somewhere there is an undefined transition. Undefined transitions on the way to the left are detected by directly comparing the sizes of the two sets, and on the way to the right the composition of functions $\big(\gamma_a^+\big)^{-1} \circ f$ is tested.
    }
        \delta_a^-\big((f, R)\big) &= 
        \begin{cases}
            (f', R'), & \text{if } |R'| = |R| \text{ and } |f(R)| = \big|\big(\gamma_a^+\big)^{-1}(f(R))\big|;
            \\
            \text{undefined}, & \text{otherwise}.
        \end{cases}
    \end{align*}

    Notice, that in every reachable state $(f, R)$ from $Q_-$ the inclusion $R \subseteq \Dom f$ takes place, which can be proved by induction.
    \begin{equation*}
        Q_- = \set{(f, R)}{R \subseteq \Dom f}
    \end{equation*}

    A claim about correctness for the second pass of the constructed automaton is formulated as follows.
    \begin{claim}
        On a string $uv$, after reading it from left to right, the constructed automaton then reads $v$ backwards completely if and only if all the transitions from the computation of $\widetilde{\mathcal{A}}$ on $uv$ applied on the symbols of the suffix $v$ are possible in the automaton $\mathcal{A}$. If $\mathcal{B}$ reads $v$ backwards completely, then afterwards it comes to a state $(f, R)$, where $R$ is a set of states visited by the automaton $\widetilde{\mathcal{A}}$ at the current symbol on the way back in its computation on the string $uv$;
        here $f$ is the behavior function of $\widetilde{\mathcal{A}}$ on $u$.
    \end{claim}
    If, after the second pass, the automaton $\mathcal{B}$ eventually comes to the left end-marker without rejecting on the way, then this means that the input string is indeed accepted by the given automaton $\mathcal{A}$. Hence, at the third pass, there is nothing left for the automaton $\mathcal{B}$ to do, except to read the string from left to right and accept in the end. Also, at this point the automaton remembers a pair $(f, R)$, where $f$ coincides with the function $\gamma_{\vdash}$, restricted to the set $Q_-$. Hence, it can reversibly cast aside the behavior function. The set $R$ is a subset of the set $\Dom \gamma_{\vdash} \setminus \{ p_0 \}$, because for every state $(f, R)$ it is true that $R \subseteq \Dom f$.
    \begin{align*}
        Q_+ &\supseteq \set{R}{R \subseteq \Dom \gamma_{\vdash} \setminus \{ p_0 \}}
        \\
        \delta_{\vdash}\big((\gamma_{\vdash}, R)\big) &= R
        \\
        \delta_a^+(R) &= R
        \\
        F &= \set{R}{R \subseteq \Dom \gamma_{\vdash} \setminus \{ p_0 \}}
    \end{align*}

    Let us calculate the state complexity of this transformation. Let the given automaton $\mathcal{A}$ have the set of states $P_+ \cup P_-$ with $|P_+| = k$ and $|P_-| = \ell$. Then,
    \begin{align*}
        |Q_+| &= k \cdot \binom{\ell}{m} \cdot \binom{k - 1}{\ell - m} \cdot (\ell - m)! + 2^{\ell - m},
        \\
        |Q_-| &= \binom{\ell}{m} \cdot \binom{k}{\ell - m} \cdot (\ell - m)! \cdot 2^{\ell - m}.
    \end{align*}
\end{proof}

\section{Hierarchy of MRFA with a bounded number of initial states}\label{section:MRFA-hierarchy}

In this section, a hierarchy of MRFA in the number of initial states is obtained. The bottom level of this hierarchy is 1RFA, that is, MRFA with one initial state. The $k$-th level contains MRFA with at most $k$ initial states (MRFA$^k$). The entire hierarchy is contained in the class of MRFA with unbounded number of initial states. And, as shown by the following example, every additional initial state increases the expressive power of the model.

\begin{example}\label{example:k-entry-MRFA-not-k-1-entry-MRFA}
    For each $k \geqslant 2$, the language $L_k = \bigcup_{i = 1}^k (ab^i)^*$ is recognized by an MRFA with $k$ initial states, but no MRFA with fewer than $k$ initial states recognizes this language.
\end{example}
\begin{proof}
    An MRFA$^k$ recognizing the language $L_k = \bigcup_{i = 1}^k (ab^i)^*$ accepts the strings from each set $(ab^i)^*$ from a separate initial state. For this, in each $i$-th initial state, there is a cycle by a single string $ab^i$.

    Suppose, for the sake of a contradiction, that there exists an MRFA $\mathcal{A}$ with at most $k - 1$ initial states, and $\mathcal{A}$ recognizes $L_k$.
    
    For each $i$, with $1 \leqslant i \leqslant k$, fix any initial state of the automaton $\mathcal{A}$ from which it accepts infinitely many strings from a set $(ab^i)^* \subset L_k$. Let $q_i$ be this state. Then, by reading a string long enough from $(ab^i)^*$, the automaton loops in $q_i$ by a string $ab^i$.

    Thus, for each $i$, there is a cycle in the initial state $q_i$ by some power of a string $ab^i$, and the state $q_i$ itself must be accepting. There are in total $k$ different indices, but $\mathcal{A}$ has at most $k - 1$ initial states. Hence, there is some initial state $q_i$ with cycles by some power of the string $ab^i$ and by another power of the string $ab^{i'}$, with $i \neq i'$. Accordingly, we obtain that from this initial state the automaton $\mathcal{A}$ accepts a string from the language $(ab^i)^+ (ab^{i'})^+$, and such a string not in $L_k$, a contradiction. Therefore, there is indeed no MRFA with at most $k - 1$ initial states for the language $L_k$.
\end{proof}

Earlier, it has been proved that the class of sRFA is included in MRFA. Now, the intermediate classes of MRFA$^k$ have been added to the hierarchy, and a question arises, how do they compare to sRFA. It is already known from Example~\ref{example:a*-cup-b*} that MRFA$^2$ are not contained in sRFA. And could sRFA be simulated by MRFA with at most $k$ initial states? The answer is given in the next example.

\begin{example}\label{example:sRFA--not-k-1-entry-MRFA}
    For each $k \geqslant 2$, the language $L_k = \bigcup_{i = 0}^{k - 1} (ab^i)^* ba^i$ is recognized by sRFA, but it is not recognized by any MRFA with fewer than $k$ initial states.
\end{example}
\begin{proof}
    Let us prove, that there exists a sweeping reversible automaton (sRFA) for $L_k$. The sRFA is constructed with acceptance on both sides, and uses a set of states $Q=Q_+ \cup Q_-$, where
    \begin{align*}
        Q_+ &= \{ q_0 \},
        \\
        Q_- &= \set{r_i}{0 \leqslant i \leqslant k - 1} \cup \set{r_{i, j}^*}{0 \leqslant j \leqslant i \leqslant k - 1}.
    \end{align*}
    
    The automaton does not do any computations on the first pass, and just moves its head all the way to the right.
    \begin{align*}\label{example:sRFA-not-k-1-entry-MRFA}
        \delta_a^+(q_0) = \delta_b^+(q_0) &= q_0,
        \intertext{%
        Next, after a turn at the right end-marker in the state $r_0$, the sRFA begins to read symbols $a$ from the end of a string, incrementing the number of its state until it comes to the rightmost symbol $b$. And then, the automaton moves by $b$ from the current state $r_i$ to the state $r^*_{i, 0}$.
        }
        \delta_{\dashv}(q_0) &= r_0
        \\
        \delta_a^-(r_i) &= r_{i + 1}
        && (0 \leqslant i \leqslant k - 2)
        \\
        \delta_b^-(r_i) &= r^*_{i, 0}
        && (0 \leqslant i \leqslant k - 1)
        \intertext{%
        Then, the automaton starts to check whether the remaining prefix of the input string is indeed from the language $(ab^i)^*$. For this, it has a cycle in states $r^*_{i,0}$, \ldots, $r^*_{i,i}$ with transitions by the symbols of the string $ab^i$.
        }
        \delta_b^-(r_{i, j}^*) &= r^*_{i, j + 1}
            && (0 \leqslant j < i \leqslant k - 1) \\
        \delta_a^-(r_{i, i}^*) &= r_{i, 0}^*
            && (0 \leqslant i \leqslant k - 1)
        \intertext{%
        And finally, the constructed sRFA accepts in states $r_{i, 0}^*$ at the left end-marker, and hence, checks that the input string begins with zero or more complete substrings $ab^i$.
        }
        F &= \set{r_{i, 0}^*}{0 \leqslant i \leqslant k - 1}
    \end{align*}
    Both functions $\delta_a^-$ and $\delta_b^-$ are injective, hence the automaton is reversible.
    
    Let us show, that there is no MRFA with at most $k - 1$ initial states for the language $L_k$. The proof is similar to the proof of Example~\ref{example:k-entry-MRFA-not-k-1-entry-MRFA}. Suppose, that there exists an MRFA with at most $k - 1$ initial states for $L_k$. For each $i$, where $0 \leqslant i \leqslant k - 1$, fix an initial state $q_i$, from which the MRFA accept infinitely many strings from the subset $(ab^i)^*ba^i$ of the language $L_k$. Then, there is a cycle by the string $ab^i$ in $q_i$.
    
    There are $k$ different values of the index $i$, and at most $k - 1$ initial states of the MRFA. Hence, there is an initial state $q_i$, from which infinitely many strings from $(ab^i)^*ba^i$ and infinitely many strings from $(ab^{i'})^*ba^{i'}$ are accepted, for different $i$ and $i'$. Hence, there is a cycle by a string $(ab^i)^k$, with $k \geqslant 1$, in $q_i$, and also some string $(ab^{i'})^{k'}ba^{i'}$ is accepted from $q_i$. Thus, starting from the initial state $q_i$, the MRFA can read a string $(ab^i)^k$ and finish its computation in $q_i$. Then, by reading the string $(ab^{i'})^{k'}ba^{i'}$, the automaton comes to an accepting state and accepts the string $(ab^i)^k(ab^{i'})^{k'}ba^{i'}$. This string is not from the language $L_k$, a contradiction. Therefore, there is indeed no MRFA with at most $k - 1$ initial states for $L_k$.
\end{proof}

\begin{theorem}\label{th:k-entry-MRFA-and-sRFA-comparison}
For each $k \geqslant 1$, one-way reversible automata with $k$ initial states (MRFA$^k$) recognize a proper subfamily of the family recognized by one-way reversible automata with $k + 1$ initial states (MRFA$^{k + 1}$). Also, MRFA$^k$ are incomparable with sRFA in the expressive power.
\end{theorem}

\section*{Acknowledgement}
The work of Maria Radionova was performed at the Saint Petersburg Leonhard Euler International Mathematical Institute and supported by the Ministry of Science and Higher Education of the Russian Federation (agreement no. 075–15–2022–287).

\end{document}